\documentclass[twocolumn, trackchanges]{aastex631}

\usepackage[nolist]{acronym}
\usepackage{subfigure}
\usepackage{longtable}
\usepackage{multirow}
\usepackage{makecell}

\begin{acronym}
	\acro{GoF}{goodness of fit}
	\acro{MLE}{maximum likelihood estimator}
	\acro{IMF}{initial mass function}
	\acro{VMP}{very metal-poor} 
	\acro{EMP}{extremely metal-poor} 
	\acro{IQR}{interquartile range} 
	\acro{K-S}{Kolmogorov–Smirnov}
	\acro{DoF}{degree of freedom}
	\acro{CDF}{cumulative distribution function}
	\acro{PDF}{probability density function}
	\acro{CCSN}{core-collapse supernova}
	\acro{MCMC}{Markov chain Monte Carlo}
	\acro{PISN}{pair-instability supernova}
	\acro{PPISN}{pulsational pair-instability supernova}
\end{acronym}
\begin{document}
	
	\title{A Modified Initial Mass Function of the First Stars with Explodability Theory under Different Enrichment Scenarios}
	
	\author[0009-0001-0604-072X]{Ruizheng Jiang}
	\affiliation{CAS Key Laboratory of Optical Astronomy, National Astronomical Observatories, Chinese Academy of Sciences\\ 
		Beĳing 100101, People’s Republic of China
	}
	\affiliation{School of Astronomy and Space Science, University of Chinese Academy of Sciences\\
		Beĳing 100049, People’s Republic of China
	}
	
	\author[0000-0002-8980-945X]{Gang Zhao}
	\affiliation{CAS Key Laboratory of Optical Astronomy, National Astronomical Observatories, Chinese Academy of Sciences\\ 
		Beĳing 100101, People’s Republic of China
	}
	\affiliation{School of Astronomy and Space Science, University of Chinese Academy of Sciences\\ 
		Beĳing 100049, People’s Republic of China
	}
	
	\author[0000-0002-0389-9264]{Haining Li}
	\affiliation{CAS Key Laboratory of Optical Astronomy, National Astronomical Observatories, Chinese Academy of Sciences\\ 
		Beĳing 100101, People’s Republic of China
	}
	
	\author[0000-0003-0663-3100]{Qianfan Xing}
	\affiliation{CAS Key Laboratory of Optical Astronomy, National Astronomical Observatories, Chinese Academy of Sciences\\ 
		Beĳing 100101, People’s Republic of China
	}
	
	\begin{abstract}
		
		The most metal-poor stars record the earliest metal enrichment triggered by Population III stars. By comparing observed abundance patterns with theoretical yields of metal-free stars, physical properties of their first star progenitors can be inferred, including zero-age main-sequence mass and explosion energy. In this work, the initial mass distribution (IMF) of first stars is obtained from the largest analysis to date of 406 very metal-poor stars with the newest LAMOST/Subaru high-resolution spectroscopic observations. However, the mass distribution fails to be consistent with the Salpeter IMF, which is also reported by previous studies. Here we modify the standard power-law function with explodability theory. The mass distribution of Population III stars could be well explained by ensuring the initial metal enrichment to originate from successful supernova explosions. Based on the modified power-law function, we suggest an extremely top-heavy or nearly flat initial mass function with a large explosion energy exponent. This indicates that supernova explodability should be considered in the earliest metal enrichment process in the Universe. 
  
  % We analyzed a large,  of 406 very metal-poor stars, and derived their corresponding initial mass function under different enrichment assumption. A Salpeter-like function fails to explain the 
  
  % Large discrepancies exist with a spike around $25\ M_\odot$ between the derived and theoretical distribution when a Salpeter-like function is adopted. Here we modified the standard power-law function with explodability theory. The derived Population III mass distribution can be well explained 
  % and successfully explained the observed mass distribution, supporting an extremely top-heavy initial mass function. 
            % with $\alpha_m=-0.24^{+0.77}_{-0.73}$. 
		
		% This work investigates the acIMF of Population Rn3 stars under both mono- and duo-enrichment assumptions. The explodability restriction is employed on the calculated metal-free supernova models to constrain their predictive remnant mass. The progenitor masses are derived under either mono- or duo enrichment assumptions, and the dilution factors conceptually demonstrate how the metal-enriched yields from different supernovae contribute to the forming gas of their descendant stars. Moreover, to restore the metal enrichment procedure in the real Universe as closely as possible, the multiplicity of a star is determined by the fitted dilsution factors.
		
	\end{abstract}
	
	% Unified Astronomy Thesaurus concepts:
	
	\section{Introduction} \label{sec:intro}
	
	As predicted by the Big Bang nucleosynthesis theory, Population III stars have formed in the early Universe, where only the elements H, He and trace amounts of Li existed. The heavy elements were first synthesised within these ancient stars and were ejected into the ambient gas by subsequent supernova explosions. Given the unavoidable fragmentation and stochastic processes involved in the primordial star formation, a wide spectrum of stellar masses is expected, spanning the substellar regime up to several hundred solar masses \citep{klessen_first_2023}. While stellar mass serves as a fundamental parameter in its evolution \citep{kippenhahn_stellar_2013}, current knowledge regarding the \ac{IMF} of the first stars remains rather limited. Nevertheless, recent numerical simulations \citep[e.g.][]{stacy_constraining_2013a, hirano_one_2014, hirano_primordial_2015, jaura_trapping_2022} have consistently yielded an approximately logarithmically flat \ac{IMF}.
 
    % Due to the absence of metal elements as cooling materials, these ancient stars are supposed to be enormously massive. This implies that Population III stars cover a wider range of masses than stars forming today, and that the \ac{IMF} of Population III stars is theoretically different from that of present-day stars. 
	
	The metal-free stars began to form approximately $100-200\ \mathrm{Myr}$ after the Big Bang \citep{tegmark_how_1997}. The majority of these stars were massive, surviving for only several to hundreds of million years according to their initial masses, and exploded as supernovae. However, there exists no direct detection of metal-free stars currently. The low-mass part hardly survive to the present day \citep{stacy_first_2014}. On the basis of available observations, an alternative indirect method is currently adopted to derive the physical properties of Population III stars of studying their descendant stars \citep[e.g.][]{kulkarni_chemical_2013, hartwig_descendants_2018, sodini_evidence_2024a, vanni_chemical_2024}. The ejecta from supernovae of Population III stars mixed with the pristine metal-free gas and brought in the initial metal enrichment of the Universe. Second-generation stars were formed in the metal-enriched gas, underwent only the first metal-enrichment phase in the early Universe and are thereby iron-deficient according to current nucleosynthesis theory. The long-lived low-mass stars with low metallicity preserve their abundance patterns until the present day. These old second-generation stars, enriched solely by the supernova explosions of the first-generation stars, shall be the \ac{EMP} and \ac{VMP} stars, typically refer to the Fe-deficient stars of $\mathrm{[Fe/H]<-3}$ and $\mathrm{[Fe/H]<-2}$ \citep{frebel_near-field_2015}. 
	
	Massive stars with masses $M$ in the range of $10\ M_\odot \lesssim M \lesssim 100\ M_\odot$ end with gravitationally-induced collapse. A successful \ac{CCSN} explodes and the synthesized elements are ejected into the ambient gas, producing a typical abundance pattern observed in \ac{EMP} stars. However, a failed supernova would collapse directly into a black hole with no explosion observed \citep{de_bennassuti_limits_2017}. In this work, the supernova explodability is considered to ensure the theoretical supernova model realistic. Stars with mass from $100\ M_\odot$ to $140\ M_\odot$ end their lives as \ac{PPISN} \citep{woosley_pulsational_2017}. The nuclear flashes are not sufficiently energetic to disrupt the entire star and a series of pulsations occur instead. Besides, stars with masses of $140\ M_\odot \lesssim M \lesssim 260\ M_\odot$ are completely disrupted by a violent explosion with no remnant left, while stars above $260\ M_\odot$ would collapse directly into black holes \citep{heger_nucleosynthetic_2002}. 
	
	Scientific interest has arisen recent years from the metal-enrichment procedure and the subsequent formation of second-generation stars. It has long been assumed to interpret the abundance signatures of \ac{EMP} stars with a single metal-free star. The \ac{IMF} of the first-generation stars has been subsequently inferred under the mono-enrichment assumption \citep{ishigaki_initial_2018}. However, \citep{cooke_carbon-enhanced_2014} suggested that a small multiple of first stars could also form these metal-poor second-generation star. Additionally, \citet{hartwig_machine_2023} suggested that only a small fraction ($31.8\%\pm2.3\%$) of \ac{EMP} stars can be classified as mono-enriched. In this research, we assume a duo-enrichment scenario, where two separate Population III supernovae together contribute to the observed abundances of an \ac{EMP} star. The duo-enrichment scenario presents a suitable representation of the mixing of multiple supernovae without introducing possible over-fitting situation caused by limited abundance detections (Section \ref{subsec:data select}). 
	
	In this work, we introduce a revised abundance fitting procedure (Section \ref{sec:method}), firstly incorporating restrictions on supernova explodability. The adopted data are selected from a large sample of 406 \ac{VMP} stars from high-resolution spectroscopic observations (Section \ref{sec:data}). The derived \ac{IMF} is present in in Section \ref{sec:result}, along with the analysis of the potential effect of observational uncertainties as well. Various distribution functions have been fitted to the \ac{IMF}, among which the explodability-modifying power-law function well explain the observed mass distribution (Section \ref{sec:discussion}). Our findings are summarized in Section \ref{sec:summary}. 
	
	\section{Method} \label{sec:method}
	
	% This work explores the multi-enrichment scenario, which suggests that the initial metal enrichment process arises from the combined contribution of multiple Population III stars. Several distinct massive Population III stars independently exploded as supernovae in the early Universe. These ancient stars synthesized metal-enriched material within and subsequently ejected it with further explosive nucleosynthesis during supernova events into the surrounding primordial gas. 
	% The timescale of the mixing procedure varies from $10$ to $100$ $\mathrm{Myr}$ depending on the properties of the halo \citep{jeon_recovery_2014}, which is substantially shorter than that of Population II star formation. Thereby we assume an efficient mixing process where complete mixing is achieved instantaneously. 
	
	\subsection{Supernova model} \label{subsec:sn yield}
	
	We utilized the nucleosynthesis yields of non-rotating massive metal-free stars calculated by \citet{heger_nucleosynthesis_2010} as \ac{CCSN} models, where both mixing and fallback effects are considered. The Population III supernova yield grids span broad parameter spaces, including progenitor mass, explosion energy, and mixing parameter. The piston's location is flexible and can be chosen at either the entropy per nucleon $s=S/N_{\mathrm{A}}k_{\mathrm{b}}=4$ or the sudden decrease of the electron mole number $Y_\mathrm{e}$. The $s=4$ model is adopted as our fiducial choice for it has a more refined grid of explosion energy. The theoretical yield predictions are assembled in the STARFIT code\footnote{publicly accessible at \url{https://starfit.org/}}, which has been effectively used in various investigations and successfully interpreted the light-element abundance patterns of some \ac{EMP} stars. 
	
	Over $140\ M_\odot$, stars are completely disrupted and expel all the mass with large metal ejecta. If a top-heavy Population III \ac{IMF} is assumed, a considerable fraction of \ac{PISN}e should be expected \citep{heger_how_2003}. Furthermore, a recent study by \citet{xing_metal-poor_2023} has found an evidence for the existence of a descendant metal-poor star of \ac{PISN}. However, there is still some debates on the fraction of \ac{PISN} contribution due to the discrepancies in abundance from different spectra \citep{thibodeaux_lamost_2024, skuladottir_pair-instability_2024, jeena_origin_2024}. Besides, the adoption of different theoretical supernova yields could also lead to a different result \citep{jeena_core-collapse_2023}. In consideration of the completeness of theoretical models for massive stars, \ac{PISN} models are still considered, for which the nucleosynthesis yields by \citet{heger_nucleosynthetic_2002} is adopted in this work.
    % Furthermore, a recent study by \citet{xing_metal-poor_2023} has found a possible evidence for the existence of a descendant metal-poor star of \ac{PISN}. Although there exists some tension in its origin \citep{jeena_core-collapse_2023, jeena_origin_2024, thibodeaux_lamost_2024}, \ac{PISN} models are still considered for the completeness of theoretical models for massive stars. In this work, we adopted the theoretical yields from \citet{heger_nucleosynthetic_2002} for \ac{PISN}.
    Stars with helium cores in the mass range $M_\mathrm{He}=65-130 M_\odot$ are calculated in these \ac{PISN} models. This mass range corresponds to main-sequence stellar masses of approximately $M=140-260 M_\odot$ by an empirical relation of 
	\begin{equation}
		M_\mathrm{He}\approx\frac{13}{24}\left(M-20\ M_\odot\right). 
	\end{equation}
	Stars end with \ac{PPISN}e lose most of elements heavier than magnesium towards their subsequent black holes \citep{woosley_pulsational_2017}. Hence \ac{PPISN} models are not considered in this work. 
	
	% The inconsistency between nucleosynthesis prediction and observational abundance suggest the probable 
	Certain elemental abundances are combined or excluded in supernova models for various reasons. Carbon and nitrogen abundances are combined, considering the influence of material mixing in the dredge-up stage. Additionally, predictions for Sc, Cr, Cu, Zn are disregarded for they have alternative nucleosynthesis contributions \citep{heger_nucleosynthesis_2010}. Both Sc and Cu are also ignored due to potential contamination from other nucleosynthesis processes. This elemental selection results in only 14 (combined) elements being included: C+N, O, Na, Mg, Al, Si, S, Ca, Ti, V, Mn, Fe, Co, Ni. Note that S abundance is only provided for one star, CS 30315-029, that satisfies our target selection criteria \citep{siqueira_mello_high-resolution_2014}. 
	
	\subsection{Abundance fitting} \label{subsec:abund fit}
	
	The abundance fitting procedure used in this work is based on the STARFIT code, an automated fitting program designed to seek the best-fitting model by $\chi^2$ minimization. The $\chi^2$ residual is defined similarly to the standard form demonstrated in \citet{heger_nucleosynthesis_2010}, 
	\begin{equation}
		\chi^2 = \sum_{i=1}^N \frac{(F_i-D_i)^2}{\sigma^2_i}. 
	\end{equation}
	Here $F_i$ and $D_i$ are the fitted and observed values of $\log\epsilon_i$\footnote{$\log\epsilon_i=\log(N_i/N_\mathrm{H})+12$, where $N$ is the number density of atoms a given element.} for an element $i$, while $\sigma_i$ is the corresponding observational uncertainty. Upper limits are excluded (see Section \ref{sec:data}). The fitted value $F_i$ represents the supernova yields diluted with metal-free gas. According to enrichment scenario, the functional form of the fitted value may change. The mixing process is influenced by different factors such as ejecta mass from supernova, mixing efficiency, and primordial gas distribution. The ejecta mass can be estimated through nucleosynthesis calculations and explosion models. A dilution factor, noted as $O$, is introduced to physically mix the supernova yields with the primordial gas \citep{tominaga_supernova_2007}. For a mono-enrichment assumption, a single dilution factor is used to simplify the mixing and enriching process. Therefore, the fitted value $F_i=M_i+O$ is equivalent to the sum of model yield value $M_i$ and the dilution factor $O$. The lower limit of dilution factor \citep{magg_minimum_2020} is not applied in this work. However, the mixing process becomes more complicated under multi-enrichment assumption. The ejected metals from neighboring supernovae mixed in the interstellar medium and shaped the formation environment of the second-generation stars. Each supernova requires its individual dilution factor to describe its independent mixing process. We treat the dilution factors as free parameters to better delineate a complex inhomogeneous mixing. The fitted elemental abundance is a combination of diluted supernova yields, 
	\begin{equation}
		F_i = \log \sum_{j=1}^K 10^{M_{i,j}+O_j}. \\
	\end{equation}
	In this equation, $j$ denotes one of the $K$-combination of different supernova models, and $O_j$ is the dilution factor for supernova model $j$. 
	
	Specifically, the abundance terms to be combined (such as C+N) are substituted with, 
	\begin{equation}
		\frac{(F_c-D_c)^2}{\sigma^2_c}, 
	\end{equation} 
	where $c$ denotes the combined term. The corresponding uncertainty $\sigma_c$ of $N_c$ combined elements satisfies: 
	\begin{equation}
		\sigma_c\sum_{i=1}^{N_c}\ 10^{D_i} = \sum_{i=1}^{N_c}\sigma_i 10^{D_i}. 
	\end{equation}
	It is worth noting that an underlying model comparison between \ac{CCSN} and \ac{PISN} is performed in the reduced chi-square minimization process. 
	
	\subsection{Explodability} \label{subsec:explod_theory}
	
	Presupernova stars with masses below $100\ M_\odot$ supposedly follow two potential evolutionary paths: they can either explode as supernovae or form into black holes, which is denote by a parameter explodability $\zeta$. The explodability parameter $\zeta$ is set to $\zeta=1$ and $0$ for explosive and non-explosive supernovae respectively. The prevalent approach to constraining the properties of the progenitor first stars with \ac{EMP} stars only provides insights into those progenitors that explode as supernovae. The original presupernova models explore a wide range of free parameter space. However, some models lead to unrealistic abundance predictions, probably due to the unusual occurrence of the corresponding parameter combinations in nature. Therefore, it is suggested that supernova explodability theory should be integrated with supernova yield models. 
	
	Several attempts from different researchers have been made to distinguish between the exploding and non-exploding cases. To assess the explodability, \citet{oconnor_black_2011} proposed that progenitors with a bounce compactness $\xi_{2.5}$ less than or equal to $0.45$ are most likely to collapse into black holes. The compactness $\xi_M$ at the time of core bounce is defined as: 
	\begin{equation}
		\xi_M = \left.\frac{M/M_\odot}{R(M)/1000\ \mathrm{km}}\right|_{t=t_\mathrm{bounce}}. 
	\end{equation}
	\citet{ertl_two-parameter_2016} argued a two-parameter criterion based on the stellar mass $M_4$ and the mass derivative $\mu_4$ at the entropy per nucleon of $s=4$. The related parameters are defined as follows: 
	\begin{eqnarray}
		M_4 &= M(s=4)/M_\odot, \\
		\mu_4 &= \left.\frac{dM/M_\odot}{dr/1000\ \mathrm{km}}\right|_{s=4}. 
	\end{eqnarray}
	A most recent study by \citet{boccioli_explosion_2023} suggested that the presence of density jumpy near the Si/Si-O interface is the sign of explosion. Although various theoretical efforts were made for a large number of progenitor models, similar discrete islands of failed explosions were found around $15$, $20$, and $25\ M_\odot$ \citep{pejcha_explosion_2020}. 
	
	Considering that the explodability calculations mentioned above are based on solar-metallicity models, we adopted an alternative criterion based on the metal-free supernova evolution calculation \citep{zhang_fallback_2008}. They suggested an upper limit on baryonic remnant mass $M_\mathrm{baryon}$, above which black holes may form. The limiting mass can be computed from the following equation: 
	\begin{equation} \label{eqn:max_remnant}
		M_\mathrm{baryon} = M_\mathrm{remnant}\left(1-\frac{3}{5}\frac{GM_\mathrm{remnant}}{R_\mathrm{remnant}c^2}\right)^{-1}. 
	\end{equation}
	Here, $M_\mathrm{remnant}$ and $R_\mathrm{remnant}$ represent the gravitational mass and the radius of the remnant respectively. By selecting a maximum gravitational mass of $1.7\ M_\odot$ or $2.0\ M_\odot$, one can infer a maximum baryonic mass of $1.94\ M_\odot$ or $2.35\ M_\odot$. It efficiently selects the optimal supernova models, resulting in a notable reduction in the total number of models from 17,640 to 10,332 or 8,974 based on the adoption of the maximum mass of neutron stars. The explodability theory helps constraint model parameter spaces, rendering the yields of nucleosynthesis more plausible. Furthermore, \citet{hartwig_machine_2023} offered an alternative observation-based method to confirm realistic model yields. 
	
	In contrast, stars with main-sequence mass in the ranges of $100-140\ M_\odot$ end with \ac{PPISN} and collapse directly into a black hole \citep{woosley_pulsational_2017}, while star with mass in $140-260\ M_\odot$ are commonly believed to explode with no remnant as \ac{PISN} \citep{heger_nucleosynthetic_2002}. Under each supernova model, explodability parameter remains constant (i.e. $\zeta_\mathrm{PPISN}=0$ and $\zeta_\mathrm{PISN}=1$) with mass changes. Hence, both \ac{PPISN} and \ac{PISN} models are not theoretically constrained by explodability parameter. 
	
	\section{Data} \label{sec:data}
	
	The primary source of abundance data is the SAGA database \citep{suda_stellar_2008}, supplemented by high-resolution spectroscopic observations of $385$ \ac{VMP} stars that are not included at the website. Upper detection limits, which inadequately constrain the physical properties of progenitor stars, are treated as non-detections. The SAGA database comprises a significant number of metal-poor stars from various papers. This study utilizes the most recent available version on the SAGA database website, covering nearly all publications up to 2019. 
	
	\subsection{Data selection} \label{subsec:data select}
	
	Data selection is based on two distinct criteria: metallicity and number of elements with abundance determination. Firstly, a conventional understanding has that Fe-deficient stars are predominantly enriched by Population III supernovae. Stars with detected metallicity of $\mathrm{[Fe/H]}<-3$ are usually considered true descendants of Population III \ac{CCSN}e \citep{frebel_near-field_2015}. For \ac{PISN} descendants, theoretical predictions \citep[e.g.][]{karlsson_uncovering_2008, de_bennassuti_limits_2017, salvadori_probing_2019, magg_metal_2022} suggested that they could be found at higher metallicities. Besides, a highly possible \ac{PISN} descendant with $\mathrm{[Fe/H]=-2.42\pm0.12}$ was identified by \citet{xing_metal-poor_2023}. Therefore, the metallicity threshold in this work is set to $\mathrm{[Fe/H]}<-2.5$. Secondly, an additional restriction is placed on the available elemental abundance number $N$ to avoid the over-fitting caused by insufficient observed data. This restriction on abundance number is based on the principle of having more observations than predictors in fitting procedure. In each instance of a supernova event, the evolution and nucleosynthesis model incorporates three parameters: the progenitor mass, the explosion energy, and the mixing parameter. An additional free-fitted parameter, the dilution factor, is required to depict the dilution by the pristine gas. Therefore, the abundance number threshold is set to $N \geq N_\mathrm{min} = \sum_{j=1}^K P+1$, where $j$ denotes one of the different supernova models in a $K$-enrichment situation. For the duo-enrichment assumption investigated in this study, the $N$ thresholds are set to $N_\mathrm{min} = 8$. 
	
	The threshold of the abundance number is validated through the following process. Reduced chi-squared residual $\chi^2_\nu = \chi^2/\nu$, where $\nu$ represents the \ac{DoF}, is commonly used to assess and compare different models. If the resulting $\chi^2_\nu$ is significantly greater than $1$ ($>5$ in this work), the model fitting is considered a bad fit, while if $\chi^2_\nu$ is substantially less than $1$ ($<0.5$ in this work), it is categorized as over-fitting. However in a non-linear model, it is challenging to determine the \ac{DoF}, which may vary between $0$ and $N-1$ \citep{andrae_dos_2010}. The \ac{DoF} is arbitrarily set at $\nu=N-P$ for simplicity. 
	
\begin{figure*}[htbp]
		\centering
		\subfigure[]{
			\label{fig:chisqr_abunum:a}
			\includegraphics[height=0.4\textwidth]{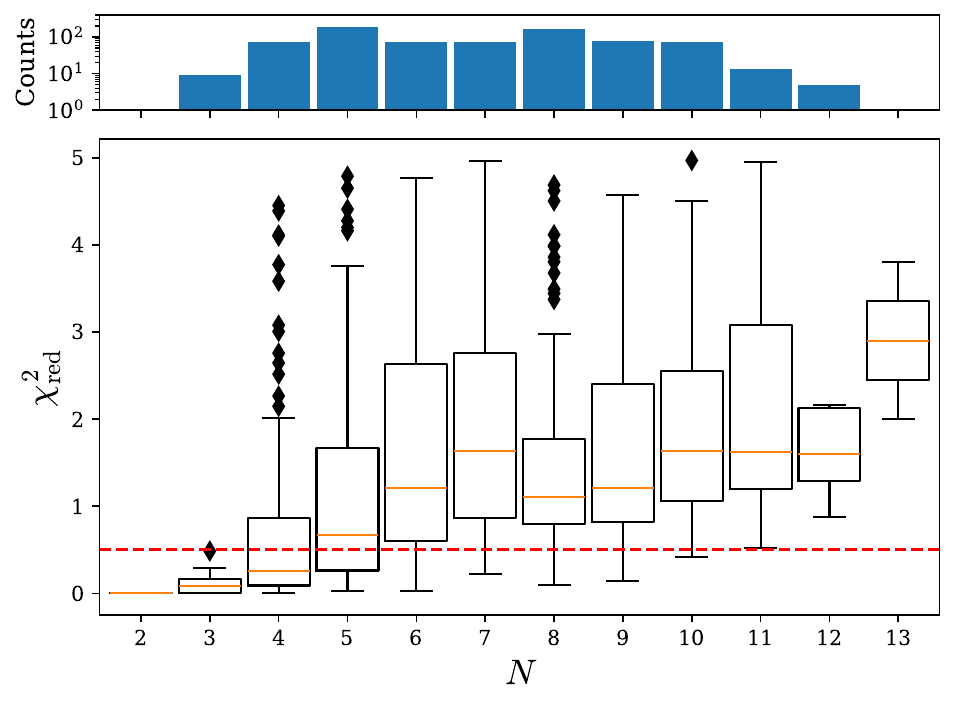}
		}
		\subfigure[]{
			\includegraphics[height=0.4\textwidth]{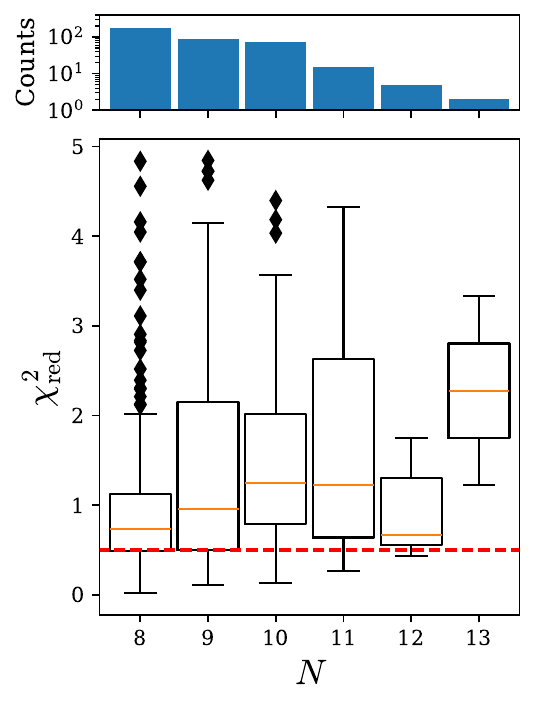}
			\label{fig:chisqr_abunum:b}
		}
		\caption{\subref{fig:chisqr_abunum:a} Comparison of the distributions of the $\chi^2_\mathrm{red}$ varying with the abundance number $N$, as well as histogram of $N$ in the upper panel. For a given $N$, the box extends from the first quartile to the third quartile of the corresponding $\chi^2_\mathrm{red}$, with an orange line indicating the median value. The whiskers extend to the furthest $\chi^2_\mathrm{red}$ lying within $1.5\times$ the \ac{IQR} from the box. Outliers are represented by thin diamonds beyond the whiskers. The horizontal red dotted line at $\chi^2_\mathrm{red}=0.5$ signifies an over-fitting threshold. The distribution of the observed abundance number is shown in the upper panel. \subref{fig:chisqr_abunum:b} Same as Panel \subref{fig:chisqr_abunum:a}, but for the selected stars under duo-enrichment assumption. }
		\label{fig:chisqr_abunum}
	\end{figure*}

	We select all \ac{EMP} stars with more than two elemental abundances observed, amounting to $900$ stars, to investigate the distributions of $\chi^2_\mathrm{red}$. A mono-enrichment assumption is considered in this $N$-threshold test. Under this assumption, all the stars support \ac{CCSN} yields, which suggests the free-fitted parameter $P=4$. The best-fit $\chi^2_\mathrm{red}$ distributions are presented in a box-and-whisker diagram (Figure \ref{fig:chisqr_abunum:a}), illustrating the locality and dispersion of data with the \ac{IQR}. The median $\chi^2_\mathrm{red}$ for $N \geq 5$ is all above the unit number of one, indicating a good fit. However, in the majority cases for $N \leq 4$ , the $\chi^2_\mathrm{red}$ dominates the over-fitting region ($<1$). It should be noted that, when $N$ drops under $4$, the abundance number smaller than \ac{DoF} is set to $1$ to avoid non-positive $\chi^2_\mathrm{red}$. 
	
	However, the ratio of $\chi^2$ residual to the abundance number $N$ substantially decreases when $N$ drops under $4$. This ratio is remarkably smaller than one, suggesting a shift into the over-fitting range. It can be concluded that the N-threshold $N\mathrm{min}=5$ is appropriate for the restriction to avoid over-fitting situation. Similarly, the abundance number threshold is set at $N_\mathrm{min}=8$ under duo-enrichment scenario, where double \ac{PISN}e situation is excluded in the model comparison procedure of abundance fitting. As demonstrated in Figure \ref{fig:chisqr_abunum:b}, the best-fit $\chi^2_\mathrm{red}$ of stars with $N\geq8$ predominantly locates far exceeding the over-fitting range, suggesting that the threshold is consistently applied in multi-enrichment. 
	
	\subsection{Repeated observations} \label{subsec:rep obs}
	
	The observations included in the SAGA database are compiled based on scholarly publications. Some metal-poor stars are observed by different researchers. Besides, 22 objects of the \ac{VMP} samples from \cite{li_four-hundred_2022} were published in previous researches. They are provided with different abundance adoptions for the same element. The preliminary data selection procedure does not preclude the repeated observations of stars that have been analyzed in multiple studies. An implemented selection is required to deal with the repeated stars. At first, the abundance data for a single star is taken from an individual publication. This helps to eliminate the systematic uncertainties caused by different abundance determination methods within a star. 
 
 	However, systematic uncertainties may still persist when comparing a specific element among stars from different investigations. As well discussed in \citet{suda_stellar_2017}, these discrepancies in the abundances arise from the diverse abundance determination methods employed in different studies, including adopted solar abundance, atmospheric models, atomic data, spectral lines and spectral resolving power. Note that the adopted solar abundance is properly considered in the database. The inconsistency in abundance could be ascribed to observational uncertainties in most cases, although it could be significant ($>1\ \mathrm{dex}$) for a few elements in some stars, like Ti as reported in \citet{suda_stellar_2017}. While these inconsistencies could be neglected because they hardly affect the resulting progenitors' mass distribution, we nonetheless remove these objects with caution. A priority parameter denoted as $\mathcal{P}$, as proposed by \citet{suda_stellar_2017}, is introduced to adopt abundances from more credible measurements. This parameter is defined as follows to automatically select the data with the highest spectral resolving power and the most recent publication: 
	\begin{equation}
		\mathcal{P} = (\mathcal{R} / 10000)^2 + (\mathcal{Y} - 2000), 
	\end{equation}
	where $\mathcal{R}$ and $\mathcal{Y}$ represent the spectral resolving power and the publication year. It is worth noting that the priority parameter is a man-made selection function and can be adjusted based on different scientific objectives. 
	
	The abundance data we adopted in this work are all cataloged in the SAGA database\footnote{provided at \url{https://sagadatabase.jp/}}. There might be certain differences between the cataloged data and the observed abundances from their original papers due to the referenced solar abundances. Besides, only 1D LTE value are used in this work to avoid the heterogeneity caused by NLTE corrections. We select the abundances by two distinct selection criterion, and remove the repeated observations as detailed above. A total of $406$ metal-poor stars are selected with at least $8$ desired elemental abundances observed. All of the used stars can be accessed via doi:\href{https://zenodo.org/doi/10.5281/zenodo.12578349}{10.5281/zenodo.12578349}, along with their metallicities, element numbers with determined abundances, and original reference papers. 
	
	\section{Result} \label{sec:result}
	
	\subsection{Effect of uncertainty} \label{subsec:perturb}
	
	In this work, the second generation of stars are assumed to be enriched by either a single supernova or two supernovae. As described in previous sections, nucleosynthesis predictions are fitted to the observational abundances under the two enrichment scenario. Although the best-fit \ac{GoF} could be used as classification basis of good/bad-fitting, in some cases the fitted progenitor properties are extremely sensitive to the adopted elemental abundances. Therefore, observational uncertainties would unavoidably induce perturbations of the fitted properties. The sensitivity problem to observational uncertainty have also been reported in previous studies \citep[e.g.][]{fraser_mass_2017, ishigaki_initial_2018}. 
	
\begin{figure*}[htbp]
		\centering
		\subfigure[]{
			\label{fig:pvalue_dist&GF_ratio:a}
			\includegraphics[height=0.4\textwidth]{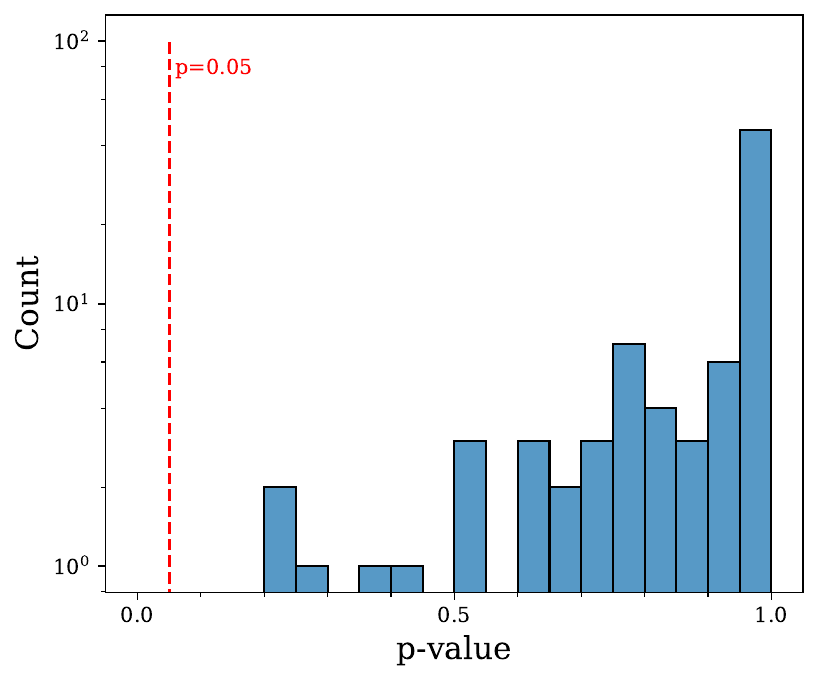}
		}
		\subfigure[]{
			\includegraphics[height=0.4\textwidth]{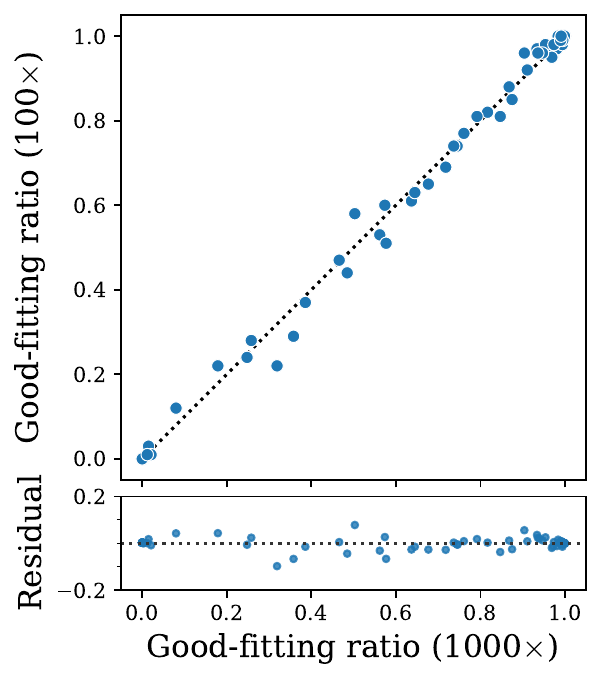}
			\label{fig:pvalue_dist&GF_ratio:b}
		}
		\caption{\subref{fig:chisqr_abunum:a} P-value distribution of \ac{K-S} tests conducted on the selected stars. The red dashed line denotes the threshold value of $a=0.05$. \subref{fig:chisqr_abunum:b} Comparison between the good-fitting ratios of $100$ and $1000$ sampling times, along with the corresponding residual. The dotted lines are reference lines. }
		\label{fig:pvalue_dist&GF_ratio}
	\end{figure*}

	Since our primary focus is on the progenitor mass, other fitted parameters, including explosion energy, mixing and dilution factor, are marginalized out in this work. The marginal distribution of the progenitor mass is analytically ambiguous due to the complex relationship between nucleosynthesis yields and progenitor properties. A Monte Carlo method is employed by simulating a Gaussian distribution for each abundance with corresponding observational uncertainty as its standard deviation. The perturbative response of fitted progenitor mass to the observational uncertainty is then examined respectively under both mono- and duo-enrichment assumptions. Under the duo-enrichment assumption, since computational cost becomes prohibitively expensive if a large sampling time is taken in the Monte Carlo method, the effect of uncertainty was tested with a small sampling time (100 iterations) for each star. A fully consistent result of progenitors is achieved for every star under such assumption, which suggests the effect of uncertainty could be neglected when the multi-enrichment situation is hypothesized. Furthermore, under the mono-enrichment assumption, we found that a relatively small sampling time (100 iterations) can also restore the mass distribution disturbed by observational uncertainties. Its validation under mono-enrichment assumption is conducted in a subsample of 82 stars whose metallicity $\mathrm{[Fe/H]<-3}$ presented in Figure \ref{fig:pvalue_dist&GF_ratio} by comparing the disturbed mass distribution at different times of $100$ and $1000$. Two aspects of consistency are studied, including the mass distribution and the proportion of good-fitting results. 
	
	First, we conducted a \ac{K-S} test to examine whether the disturbed results follow the same distribution regardless of the sampling times. P-value is calculated for each case with a predefined significant level at $a=0.05$ below which the null hypothesis will be rejected. P-values substantially exceed the significant level as presented in Figure \ref{fig:pvalue_dist&GF_ratio:a}. Second, the good-fitting ratios of different sampling times are compared and dotted in the Figure \ref{fig:pvalue_dist&GF_ratio:b}. These dots align well with the diagonal line of this figure with a maximum residual of the good-fitting ratio difference approximating to $0.1$ dex. This suggests a good consistency in the proportion of good-fitting results. The best-fit $\chi^2_\mathrm{red}$ upper limit of defining the good-fitting result does not affect the conclusion of consistency. It can be inferred that a rather limited sampling time will obtain numerical result of uncertainty's perturbation. The subsequent results are based on a Monte Carlo method of $100$ random samples. 
	
	% The perturbation behaves differently under different assumptions. For a specific star, the fitted result to the observational abundances is considered as fiducial. 
	
\begin{figure*}[htbp]
		\centering
		\includegraphics[width=.9\textwidth]{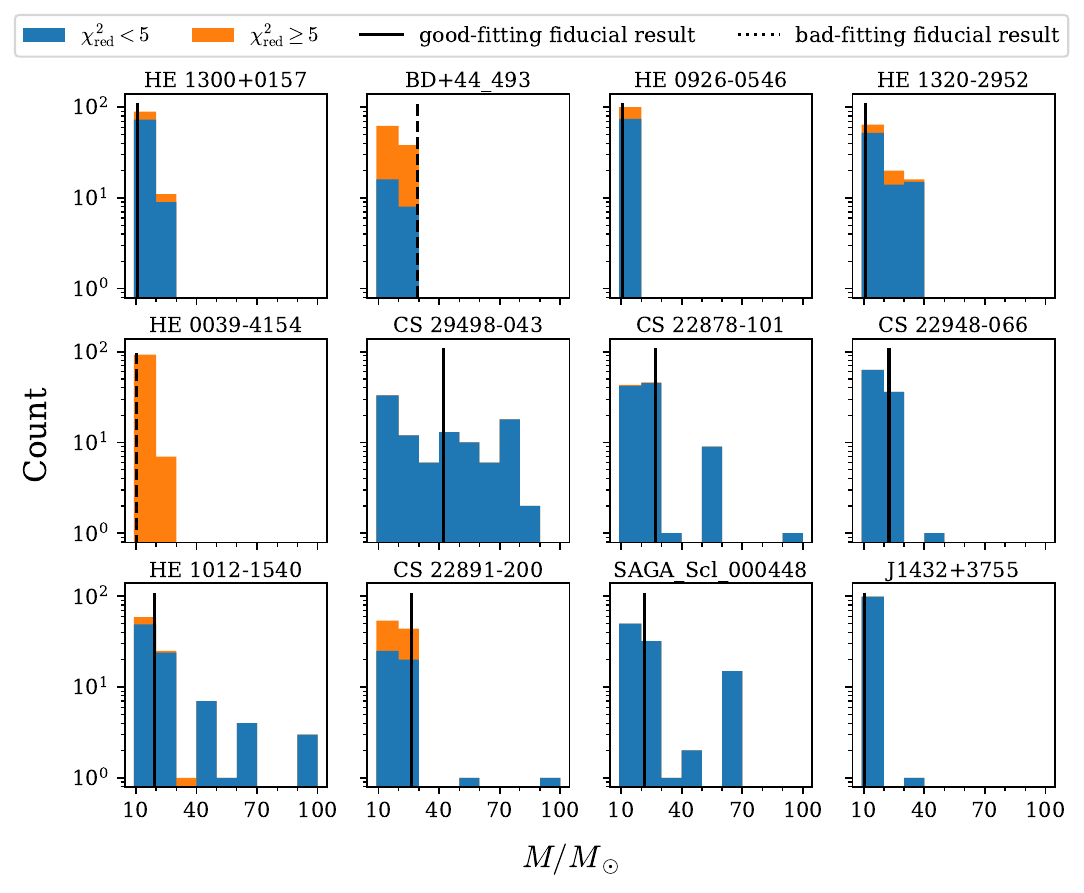}
		\caption{Some examples of the resulting disturbed mass distributions for $100$ iterations. Twelve stars are randomly selected from the adopted samples. The hatched histograms represent the fitting results, with good/poor-fittings colored in blue/orange respectively. The solid/dotted black lines denote the fiducial results of the observed abundances for good/poor-fitting. }
		\label{fig:pert_res_examples}
	\end{figure*}

	We select twelve random stars to illustrate the disturbed distribution of fitted mass under mono-enrichment assumption. As shown in Figure \ref{fig:pert_res_examples}, the disturbed mass could be perturbed to a large distance from the fiducial result. For instance in Figure \ref{fig:pert_res_examples}, the disturbed results of the star CS 29498-043 range from $10$ to $90$ solar masses, covering almost the entire mass range. Moreover, an original good-fitting fiducial result, such as BD+44 493, shifts into poor-fitting zone, which indicates that our evaluation of abundance fitting might also change due to observational uncertainties. 
	
	However when switching to the duo-enrichment scenario, the fitted results remain consistent regardless of the elemental abundances sampled by Gaussian distribution. Considering that the number of fitted parameters increase due to an additional supernova under multi-enrichment assumption, the $\chi^2$ residual of abundance fitting is bound to be improved in sacrifice of \ac{DoF}. Aimed to evaluate the \ac{DoF}-varying fitting procedure, a modified $\chi^2_\mathrm{red}$ is used for the model evaluation as defined in Section \ref{subsec:data select}. Good-fitting criterion is still set at $\chi^2_\mathrm{red}<5$ for consistency. 
	
	\subsection{Mass distribution}
	
	As discussed above, the fitting results behave quite differently under different enrichment scenarios. In order to obtain a plausible mass distribution, we developed different selection criteria on resulting masses under different situations. 
	
	For mono-enrichment assumption, elemental abundances are highly sensitive to progenitor properties and necleosynthesis yields \citep{frebel_near-field_2015} and vice versa. In other words, the fitting results would be affected by observational uncertainties. Accordingly, high-precision observations are required to obtain robust results. Yet even with the current high-resolution spectral data, the robustness of the results is still not fully guaranteed. A Monte Carlo method is adopted to characterize the effect of uncertainties on the derived progenitor mass. Although the progenitor mass for a certain star might vary drastically under the influence of uncertainties, all of the good-fitting results from the Monte Carlo method are possible true progenitors of the observed stars. Therefore instead of a fiducial progenitor mass, a disturbed mass distribution is adopted under mono-enrichment scenario. Considering that the good-fitting disturbed distribution can be hardly expressed analytically, we randomly resampled the good-fitting results to ensure that every star is assigned to the same weight in the mass distribution. Since the number of unique stellar masses in the results of each star is less than 16 (15.9 in specific) on average, the resampling time is set to thirty in order not to alter the disturbed distribution fundamentally. As a result, thirty disturbed masses are adopted as a progenitor distribution of a certain star. 
	
	Under duo-enrichment situation, a second-generation \ac{EMP} star by definition is metal-enriched by two distinct progenitor Population III stars. The metal contributions from these progenitors are different, described by dilution factors. A progenitor with a smaller dilution factor means that the metal contribution from its corresponding supernova is proportionally smaller. If the contribution from one of the resulting progenitors is negligibly small for a certain star, this might be a supporting evidence that the star experienced only a single metal enrichment process. However, there is currently no definitive method to determine the multiplicity of a star. And considering that the multiplicity determination is not the main subject of this paper, these two progenitors are treated equally no matter how little the metal contributions are. Furthermore, \citet{hartwig_machine_2023} found that multi-enriched stars are more clustered than mono-enriched stars in the abundance spaces. Further conclusion can be given that the combined yields of two supernovae would be no more extreme than those of a single supernova. Hence the mono-enriched \ac{CCSN} metallicity upper limit ($\mathrm{[Fe/H]<-3}$) should be effective in the combined \ac{CCSN}e case as well. Since higher metallicities of \ac{PISN} descendants are suggested by previous studies \citep[e.g.][]{magg_metal_2022, xing_metal-poor_2023}, the metallicity upper limit for two \ac{PISN}e or a \ac{CCSN} and \ac{PISN} is set to a higher metallicity of $\mathrm{[Fe/H]<-2.5}$. This means that the metal-poor stars must have at least one \ac{PISN} component if its metallicity satisfies $\mathrm{-3\leq[Fe/H]<-2.5}$. 
	
	\section{Discussion} \label{sec:discussion}
	
\begin{figure*}[htbp]
		\centering
		\subfigure[]{
			\includegraphics[width=.9\textwidth]{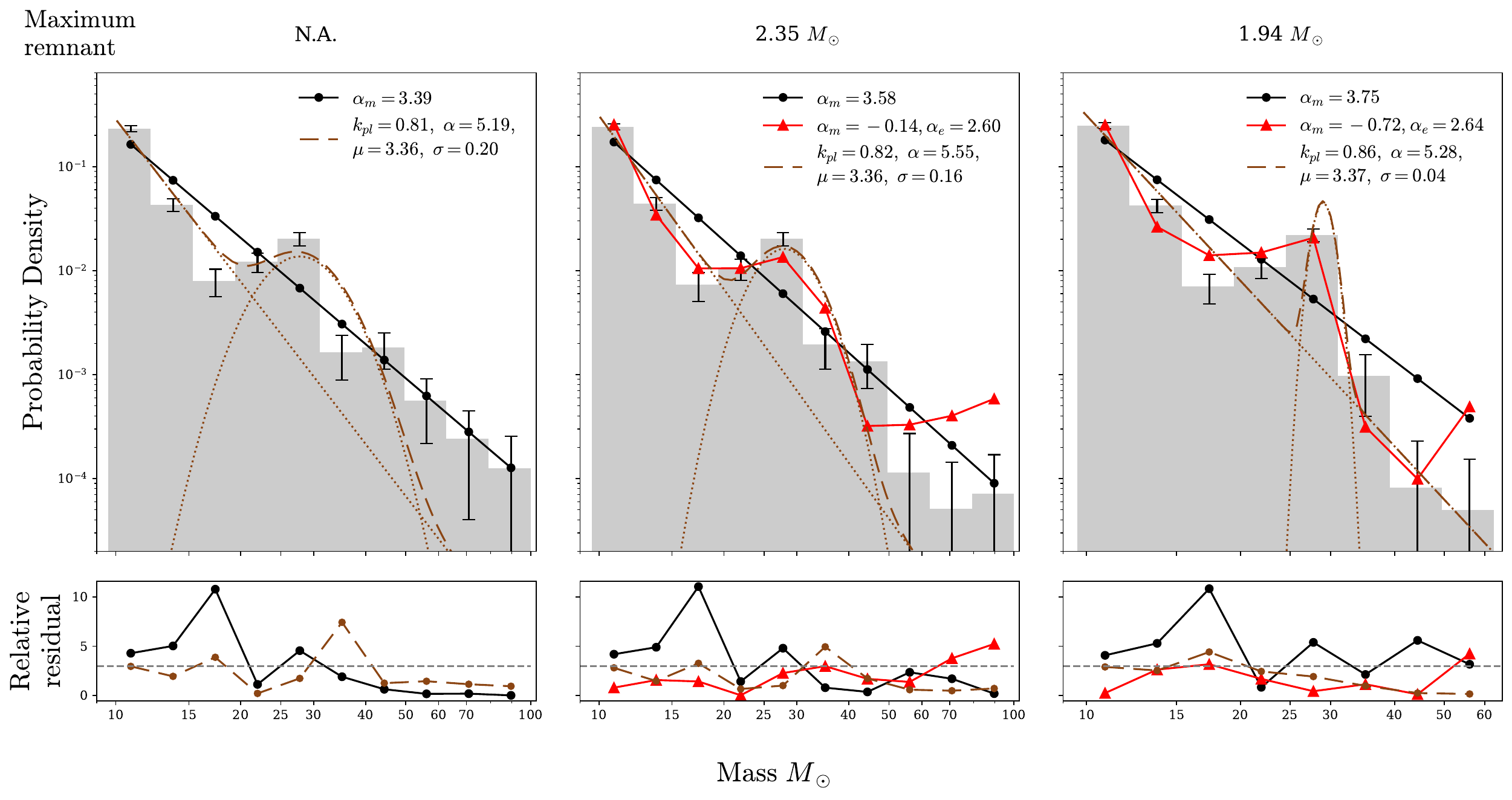}
			\label{fig:compare_dist:a}
		}
		\subfigure[]{
			\includegraphics[width=.9\textwidth]{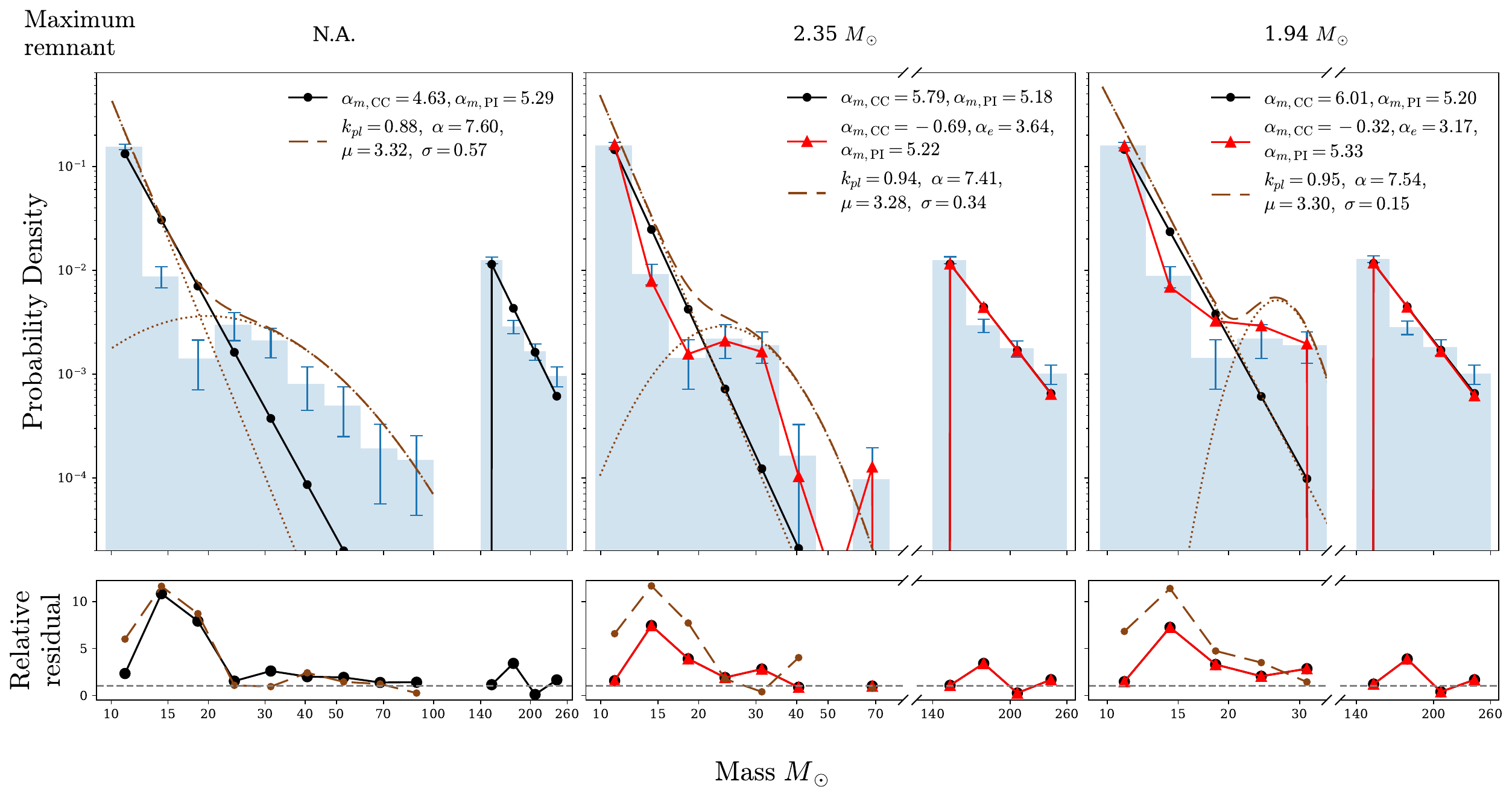}
			\label{fig:compare_dist:b}
		}		
		\caption{Inferred progenitor mass distributions under \subref{fig:chisqr_abunum:a} mono-enrichment and \subref{fig:chisqr_abunum:b} duo-enrichment assumption, along with their best-fit \ac{IMF}s and relative residuals based on different explodability constraints. The black circles with solid lines, red triangles with solid lines and brown dashed lines represent the best-fit standard power-law, explodability-modifying power-law distributions and combined log-normal power-law respectively. For detailed definition of parameters in legends, please refer to Section \ref{subsec:std_pw}-\ref{subsubsec:explod_modify_dist}. }
		\label{fig:compare_dist}
	\end{figure*}

	Since how to determine the multiplicity of the first stars remains inconclusive, it is of interest to analyze the fitting results under different enrichment assumptions based on current \ac{EMP} observations. Based on the selection criterion above, the progenitor mass distribution can be easily obtained. As shown with the colored histograms in Figure \ref{fig:compare_dist}, the derived mass probability density decreases with increasing mass on a large scale. 
    % However, the monotonic decrease of mass distribution is destroyed after $\sim15\ M_\odot$. 
    As shown in Figure \ref{fig:chisqr_abunum:a}, no \ac{PISN} progenitor is suggested in the adopted \ac{EMP} stars when mono-enrichment is assumed. Besides, an obvious spike exists around $25-30\ M_\odot$ no matter whether supernova models are constrained by explodability theory. When \ac{CCSN}'s explodability is considered, the mass distribution drops quickly around $40$ $M_\odot$. 
	
	Considering that metal-poor stars are possibly multi-enriched, the metal-poor stars are also fitted to a multi-enrichment scenario. Under this assumption, \ac{PISN} components are inferred from abundance data. As shown with the light columns in Figure \ref{fig:chisqr_abunum:b}, the \ac{PISN} component simply follows power-law function, while the spike around $25\ M_\odot$ still exists in the \ac{CCSN} mass range. 
	
	% \begin{figure*}[htbp]
	% 	\centering
	% 	\includegraphics[width=.9\textwidth]{massplot.pdf}
	% 	\caption{Derived mass distributions of Population III stars of different enrichment assumptions and maximum remnant restrictions. Light- and dark-colored histograms represents the mass distributions under multi- and mono-enrichment assumption. }
	% 	\label{fig:massplot}
	% \end{figure*}
	
	% In order to obtain the representation of the observed mass distribution, the \ac{IMF} of Population III stars have been explored by various researches. 
 	Previous researches have explored various representations of the \ac{IMF} of Population III \citep{fraser_mass_2017, ishigaki_initial_2018}. We therefore used several different distribution functions to describe the mass distribution. We also calculated their corresponding relative residuals by $|P_{obs}-P_{theory}|/\Delta P_{obs}$ to evaluate the theoretical mass functions. A Poisson observed mass distribution uncertainty $\Delta P_{obs}=\sqrt{P_{obs}/dM}$ is assumed. The best-fit \ac{IMF}s and parameter values of different functional forms are presented in Figure \ref{fig:compare_dist} and Appendix \ref{sec:param_value}. We first applied a Salpeter-like \ac{IMF}, as shown with black circles and solid lines in Figure \ref{fig:compare_dist}, with a free-fitted parameter $\alpha_m$ to represent the resulting mass distribution. A log-normal function is linearly combined with the Salpeter-like \ac{IMF} as presented with brown dashed lines in order to fit the local maxima around $25\ M_\odot$. Besides, explodability theory is also considered to explain the deviation of the resulting distribution from the power-law function as represented by the red solid lines with triangles. 
	
	\subsection{Standard power-law distribution}
         \label{subsec:std_pw}
	
	The power-law distribution is the most widely used functional form for the mass distribution. The standard power-law function with a mass exponent $\alpha_m$ follows \citet{salpeter_luminosity_1955}
	\begin{equation}
		\frac{dN}{dM} = P_M \propto M^{-\alpha_m}, 
		\label{eqn:raw_powerlaw}
	\end{equation}
	where $M_\mathrm{min}$ is the minimum of mass for which the distribution establishes. For $\alpha_m>1$,the power-law distribution can be converted into the Pareto form by viewing the minimum mass $M \geq M_\mathrm{min}$ as the scale parameter: 
	\begin{equation}
		P_{M,pl} = \frac{\alpha_m-1}{M_\mathrm{min}}\left(\frac{M}{M_\mathrm{min}}\right)^{-\alpha_m}, 
	\end{equation}
	where the prefactor $\left(\alpha_m-1\right)/{M_\mathrm{min}}$ represents the normalizing constant. When $\alpha_m>1$, the method of \ac{MLE} is often used to determine the fitted exponent $\alpha_m$ of the standard power-law distribution via 
	\begin{equation}
		\hat{\alpha}_m = 1 + n \left[\sum^n_{i=1} \ln \frac{M_i}{M_\mathrm{min}} \right]^{-1},  
	\end{equation}
	where $n$ represents the data point. And its standard deviation can be given as
	\begin{equation}
		\Delta\hat{\alpha}_m = \frac{\hat{\alpha}_m - 1}{\sqrt{n}}. 
	\end{equation}
	
	When the domain of definition for the exponent $\alpha_m$ is extended to $\alpha_m<1$, the mass minimum $M_\mathrm{min}$ and maximum $ M_\mathrm{max}$ are both needed to avoid infinite area, thus leading Equation \ref{eqn:raw_powerlaw} to the form of 
	\begin{equation}
		P_{M,pl} = \frac{M_\mathrm{min}^{1-\alpha_m}}{M_\mathrm{max}^{1-\alpha_m}-M_\mathrm{min}^{1-\alpha_m}}\frac{1-\alpha_m}{M_\mathrm{min}}\left(\frac{M}{M_\mathrm{min}}\right)^{-\alpha_m}. 
	\end{equation}
	
	The minimum and maximum mass are set spontaneously according to the parameter mass range of supernova models. As defined, the mass range for \ac{CCSN} is set at $M_\mathrm{min}=9.6\ M_\odot$ and $M_\mathrm{max}=100\ M_\odot$, while \ac{PISN} it is set for at $M_\mathrm{min}=140\ M_\odot$ and $M_\mathrm{max}=260\ M_\odot$. The mass exponents over different mass ranges, are supposed to be different to reflect different formation and evolution histories of different types of massive stars. Hence, the mass distribution over all mass range can be written as, 
	\begin{eqnarray}
		\log P_M \propto 
		\left\{
		\begin{array}{rcl}
			-\alpha_{m,\mathrm{CC}} \quad 9.6\ M_\odot-100\ M_\odot\\
			-\alpha_{m,\mathrm{PI}} \quad 140\ M_\odot-260\ M_\odot  
		\end{array}
		\right.
	\end{eqnarray}
	
	When a larger exponent range is assumed ($\alpha_m\neq1$), a \ac{MCMC} method with Bayesian model is used to obtain the posterior of $\alpha_m$ as a complement to the \ac{MLE}. Note that, the \ac{MLE} is mathematically equivalent to Maximum A Posteriori estimation with uniform prior distributions. 
	% To avoid the influence of different adoption of the bin size, the discrepancy is computed by subtracting the best-fit power-law distribution from the mass distribution derived from the observed \ac{EMP}s. The \ac{CDF} residual by definition begins at zero and ends at unity of one within the mass range. 
	
	The progenitor mass distribution based on supernova models is shown in the left panel of Figure \ref{fig:compare_dist}. The best-fit power-law distribution is represented as the black solid lines with dots. When model's explodability is not constrained, under mono- and duo-enrichment assumption, the best-fit exponent are $\alpha_m=3.38\pm0.04$ and $4.10^{+0.26}_{-0.24}$ respectively. The best-fit exponents for \ac{CCSN} generally follows a Salpeter-like IMF. However, significant discrepancies in these distributions around $15$ and $25\ M_\odot$ cannot be well explained by a standard power-law function. 
	
	\subsection{Combined log-normal power-law distribution} \label{subsubsec:combine_lnpw_dist}
	
	Our result suggests that the observed mass distributions under both enrichment scenarios share a similar power-law trend as represented by the black solid lines in Figure \ref{fig:compare_dist}. However, large discrepancies from $10\ M_\odot$ to $30\ M_\odot$ indicate that the standard function fails to explain all observed features in distributions. A combination of power-law and log-normal functions is developed as Equation \ref{eqn:combine_dist}. This incompatibility has been also reported in several \ac{IMF} researches. For example, \citet{fraser_mass_2017} estimated the existence of a shallower slope of \ac{CDF} in the approximate range of $15$ to $30\ M_\odot$, which roughly aligns with our mass range in this work. Besides, \citet{ishigaki_initial_2018} argued that the peak at $25\ M_\odot$ in mass distribution is inconsistent with the Salpeter \ac{IMF} with $\alpha_m=2.35$. Furthermore, a log-normal-like function is used to represent the mass distribution, which is defined as follows:
	\begin{equation}
		P_{M,ln} = \frac{1}{M\sigma\sqrt{2\pi}} \exp\left(-\frac{\left(\ln M - \mu\right)^2}{2\sigma^2}\right). 
	\end{equation} 
	
	As shown in Figure \ref{fig:compare_dist:a}, when a mono-enrichment scenario is assumed, the derived mass distributions in general follow a power-law function, and decay quickly after $35\ M_\odot$. However, an unexpected spike around $25\ M_\odot$ is inconsistent with the Salpeter-like function, which is similar to the peak at $25\ M_\odot$ reported by \citet{ishigaki_initial_2018}. We hence suppose a linear combination of power-law and log-normal functions with coefficients $k_{pl}$ and $k_{ln}$ ($k_{pl}+k_{ln}=1$):
	\begin{equation}
		P_M = k_{pl} \cdot P_{M,pl}(\alpha_m) + k_{ln} \cdot P_{M,ln}(\mu, \sigma). 
		\label{eqn:combine_dist}
	\end{equation}
    This combined log-normal power-law distribution is only fitted to the \ac{CCSN} part, for the log-normal distribution is aimed to represent the $25\ M_\odot$ spike. Hence the distribution of \ac{PISN} progenitors is still fitted with an independent power-law function. The best-fit combined functions are represented by brown dashed lines in Figure \ref{fig:compare_dist}. The relative residuals of combined log-normal power-law function are reduced from $10\ M_\odot$ to $30\ M_\odot$ compared to the ones of standard power-law function. Although the progenitors' mass distributions can be well fitted with the combined log-normal power-law distribution when mono-enrichment assumption is assumed, the combined function is a morphological fitting and does not provide a physical explanation for the local maxima around $25\ M_\odot$. For this reason, the mass exponent $\alpha$ of the combined function is not suitable to determine whether the \ac{IMF} is top- or light-heavy. In addition, the combined function fails to interpret the mass distribution of multi-enrichment scenario. The relative residuals remain much higher above the unit of one. 
	
	\subsection{Modifying distribution with explodability} \label{subsubsec:explod_modify_dist}
	
\begin{figure*}[htbp]
		\centering
		\includegraphics[width=.9\textwidth]{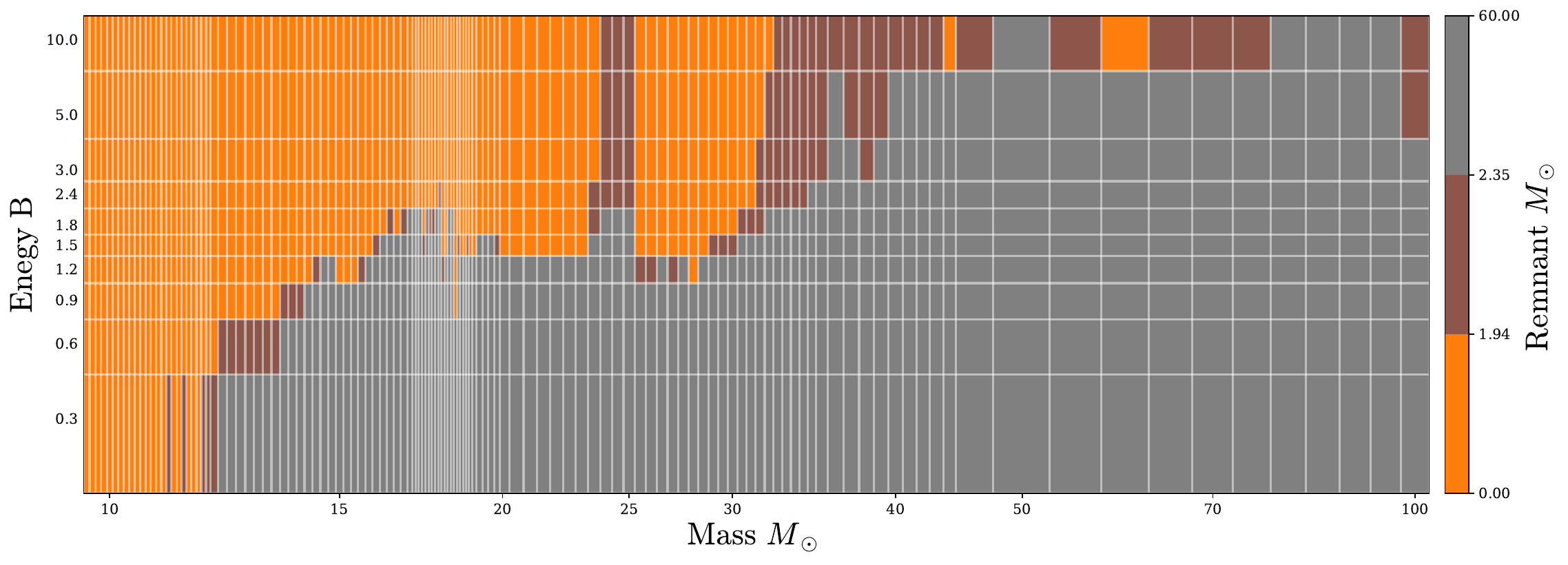}
		\caption{Baryonic remnant masses of different supernova models. The remnant maximum can be chosen between $1.94\ M_\odot$ and $2.35\ M_\odot$. }
		\label{fig:model_grid}
	\end{figure*}

	The observed distributions can be well fitted with the combined functions, as described with Equation \ref{eqn:combine_dist}. However, the linearly-combined-function fitting is basically morphological. The physical explanation behind the linear combination is not clear. It can be found that the spikes of discrepancies around $15$ and $25\ M_\odot$ in Figure \ref{fig:compare_dist} align precisely with the islands of explodability shown in Figure \ref{fig:model_grid}. Thus we attempted to explain the observed discrepancies by modifying the standard power-law distribution with explodability theory. 
	
	As commonly believed, Population III stars enriched the primordial gas in the early Universe through supernova explosions. However, for stars evolving into failed supernovae or direct collapse black holes, there exist no other channels through which the heavy elements synthesized within the stars can be rapidly fed back into the ambient gas. Therefore, the \ac{EMP} stars observed today are highly possible descendants of Population III stars that have successfully exploded as supernovae. Therefore, an explodability restriction is placed to ensure reliable supernova models. Various methods of predicting explodability have been developed as outlined in Section \ref{subsec:explod_theory}. Our analysis adopted the method of estimating a supernova's explodability by its baryonic remnant mass maximum, which is derived from the assumed theoretical upper limit mass of neutron stars. The mass maximum of neutron stars can be chosen between $1.7\ M_\odot$ and $2.0\ M_\odot$, leading to different baryonic remnant maxima of $1.94\ M_\odot$ and $2.35\ M_\odot$ respectively. According to the theoretical explodability prediction, explosion energy is needed to predict a successful or failed supernova, while mixing parameter is neglected due to its irrelevance to the supernova remnants. An independent energy distribution is thus assumed. And the modified mass function $\widetilde P_M$ can be marginalized from the bivariate probability distribution $P_{M,E}$ as 
	\begin{eqnarray}
		\widetilde P_M &=& \int \zeta\left(M,E\right)P_{M,E}dE \\ 
		&=& P_M\int \zeta\left(M,E\right)P_EdE, 
	\end{eqnarray}
	where the explodability parameter $\zeta\left(M,E\right)$ takes values from either zero or one. Assuming that both the progenitor mass and explosion energy follow the power-law function, we can simplify the modified distribution as follows, 
	\begin{equation}\label{eqn:mod_dist_simple}
		\widetilde P_M \propto M^{-\alpha_m}\int \zeta\left(M,E\right) E^{-\alpha_e} dE. 
	\end{equation}
	Given the explodability for \ac{PISN} $\zeta_\mathrm{PISN}(M, E)\equiv1$ in theoretical models, the integral term on the right side of Eq. \ref{eqn:mod_dist_simple} is constant. Because of this, the explodability-modifying mass distribution for \ac{PISN} could be converted to a simple power-law distribution. Consequently, the notation of $\alpha_e$ represents solely the energy exponent of \ac{CCSN} in this work. 
	
	This restriction leads to a limitation on the parameter spaces of the supernova models, excluding mainly the \ac{CCSN}e with massive zero-age main-sequence mass. The detailed parameter grids and corresponding remnant masses are delineated in Figure \ref{fig:model_grid}. Progenitor's main-sequence mass is thus constrained. From the supernova remnants calculated in Figure \ref{fig:model_grid}, the main-sequence masses of successfully explosive \ac{CCSN} are predicted mainly below $\sim 30\ M_\odot$. Besides, supernova models with masses around $25\ M_\odot$ in addition are excluded if a stricter remnant maximum of $1.94\ M_\odot$ is adopted. It indicates that there should exist a decrease around $25\ M_\odot$ in the mass distribution, which is consistent with the observed features of the mass distributions in Figure \ref{fig:compare_dist}. 
	
	The explodability theory excludes unrealistic supernova models before the abundance fitting procedure, which suggests that the remnant constraint is highly coupled with the fitting procedure. The mass distributions derived based on different remnant constraints are fitted with corresponding modified power-law distributions. The relative residuals are accordingly computed. All of the discrepancies substantially decline below $1$ compared to the standard power-law fitting, especially with a stricter explodability constraint. These decreases in residuals suggest that the explodability-modifying function we used can successfully explain the derived Population III mass distributions. 
	
	If a standard power-law function is fitted, the mass exponent $\alpha_{m,\mathrm{CC}}$ is increased when explodability constraint is assumed. It is mainly because that the massive main-sequence Population III stars theoretically fail to explode according to current supernova explosion theory. However, the modified power-law function recommends an extreme top-heavy \ac{IMF} with decreased mass exponent $\alpha_{m,\mathrm{CC}}$. The best-fit $\alpha_{m,\mathrm{CC}}$ under mono-enrichment decreases even below $0$, indicating that massive stars take up a larger portion than small stars. 
	Furthermore, relative residuals of duo-enrichment assumption are much smaller than those of mono-enrichment assumption. This feature in residuals indicates that the resulting mass distribution under multi-enrichment scenarios (Figure \ref{fig:compare_dist:b}) should be of more realistic. A considerable fraction of the \ac{EMP} stars should undergo multiple metal enrichment as predicted by \citet{hartwig_machine_2023}. As a result, when the maximum supernova remnant is set to $1.94\ M_\odot$ ($2.35\ M_\odot$), the best-fit mass distribution parameters are $\alpha_{m,\mathrm{CC}}=-0.32\pm0.74$ ($-0.69\pm{0.74}$), $\alpha_{e}=3.17\pm{0.30}$ ($3.64^{+0.37}_{-0.35}$) and $\alpha_{m,\mathrm{PI}}=5.33^{+0.59}_{-0.56}$ ($5.22^{+0.56}_{-0.53}$). 
	%The explodability theory with a stricter maximum supernova remnant of $1.94\ M_\odot$ matches the observed progenitor mass distribution better with a significantly lower relative residuals. The best-fit mass exponents under mono- and duo-enrichment scenarios are $\alpha_{m,\mathrm{CC}}=-1.16^{+0.09}_{-0.10}$ and $\alpha_{m,\mathrm{CC}}=-1.15\pm{0.77}$, supporting an extremely top-heavy \ac{IMF} of Population III stars. Conversely, comparatively larger energy exponents are recommended, which are $\alpha_{e}=-3.98^{+0.08}_{-0.07}$ and $\alpha_{e}=5.53^{+0.65}_{-0.57}$ respectively for different enrichment assumptions. The combination of a small mass exponent and a large energy exponent explains the abrupt peak around 25 $M_\odot$ in the progenitors' mass distribution. 
	
	Our finding suggests an extremely top-heavy (or nearly flat) Population III \ac{IMF} with a relatively large energy exponent could explain the progenitor mass distribution based on current \ac{EMP} observations. The metal-free environment in the early Universe would result in a dominance of massive stars. In the absence of cooling materials, the temperatures of star forming clouds are expected to be higher, leading to a fragmentation suppression \citep{hennebelle_physical_2024}. Besides, there are an insufficient number of strong lines in zero-metallicity stars and the adiabatic cooling dominates, leading to weak or nonexistent stellar wind \citep{krticka_weak_2010}. However, the mass exponent for the \ac{PISN} progenitors is comparatively large. It is noteworthy that the ratio between \ac{CCSN}e and \ac{PISN}e in the derived probability density function does not represent the actual ratio between explosion incidents of different supernova types in the early Universe. Since the explosion energy and metal ejecta of a \ac{PISN} are considerably larger than those of a \ac{CCSN} \citep{heger_how_2003}, vaster forming regions of second-generation stars are contaminated by \ac{PISN} explosions. More Population III descendants would have chemical component from a single \ac{PISN} than a single \ac{CCSN}. When we derive the mass distribution of the first stars based on the observations of \ac{EMP} stars, the occurrence of \ac{PISN} would be correspondingly increased. The probability of \ac{PISN} incident is thus overestimated. Further simulations are required to quantify the correction for this overestimation, specifically focusing on the detailed mixing process of gas following multiple supernova events in the early Universe.
	
	\section{Summary} \label{sec:summary}
	
	In this work, a large sample of 406 metal-poor stars are selected from the literature with high-resolution spectroscopic observations. Current observational evidence suggests that the abundances of metal-poor stars possibly originate from both \ac{CCSN}e and \ac{PISN}e \citep[e.g.][]{frebel_near-field_2015, xing_metal-poor_2023}. Therefore, different supernova models, including both \ac{CCSN} and \ac{PISN}, are used in this paper, which cover a broad range of progenitor masses from $10$ $M_\odot$ to $260$ $M_\odot$. We used a proven $\chi^2$-minimization method for the abundance pattern match between observations and theoretical predictions. We additionally set an observed element number threshold $N>\sum_{K} P$ in the abundance fitting process where $K$ represents the multiplicity. It will avoid the over-fitting cases to a large extent. 
	
	The effect of observational uncertainty on the best-fit result is analyzed under different enrichment scenarios, where the fitting result is categorized as good-fitting if its best-fit \ac{GoF} satisfies $\chi^2_\mathrm{red}<5$. Under mono-enrichment assumption, the progenitor mass can be greatly influenced by observational uncertainties. While under duo-enrichment assumption, the result are much more stable and robust, the $\chi^2_\mathrm{red}$ still remains in a suitable region and does not fall into over-fitting situation. We hence treated the resulting progenitor masses differently according to the assumed enrichment scenario. The mass distribution disturbed by observational uncertainty is adopted in the mono-enrichment scenario. Meanwhile in the case of multiple enrichment, we supposed the best-fit results to be the \ac{EMP} stars' progenitors. 
	
	Various mass distribution functions have been explored for the \ac{IMF} of the first stars in different investigations. First, a Salpeter-like distribution (i.e. standard power-law distribution) is fitted to the resulting mass distribution. However, large differences in relative residuals are found from $10\ M_\odot$ to $30\ M_\odot$. This discrepancy indicates a possible mass disappearance from $15$ to $25$ $M_\odot$ in Population III progenitors, which aligns with the islands of failed explosion predicted by various explodability researches \citep{pejcha_explosion_2020}. However, it should be noted that most supernova explodability experiments are based on solar metallicity. The predictive explodability for a certain stellar mass may vary with the metallicity, especially for the extremely metal-free progenitors. 
	
	We applied the explodability theory to explain these large relative residuals in the resulting mass distribution. Explodability constraints in addition ensure a realistic supernova model and make the nucleosynthesis yields more credible. We further modified the standard power-law function with explodability theory by adding an extra independent energy distribution. It is commonly believed that \ac{PISN} is bound to explode, the corresponding modified distribution for \ac{PISN} degenerates into the standard power-law distribution. The relative residuals of modified power-law function are drastically reduced. When the maximum mass of neutron stars is chosen as $1.7\ M_\odot$ ($2.0\ M_\odot$), an extremely top-heavy \ac{IMF} of \ac{CCSN} progenitors with $\alpha_{m,\mathrm{CC}}=-0.32$ ($-0.69$) is thus inferred, while the mass exponent for \ac{PISN} progenitors is relatively large, $\alpha_{m,\mathrm{PI}}=5.33$ ($5.22$). This modified power-law distribution offers an insight into refining and validating theoretical explodability predictions with high-resolution spectroscopic data. We highly recommend that the explodability theory should be taken into consideration in deriving the \ac{IMF} of Population III from the \ac{EMP} stars. 
	
	% \underline{Further refinements}
	This Population III \ac{IMF} investigation is based on the elemental abundance pattern sample of metal-poor stars with metallicity $\mathrm{[Fe/H]}<-2.5$. These observed metal-poor stars in the Milky Way and its satellite dwarf galaxies help demonstrate the earliest star-formation history in the Universe and provide a glimpse into the nature of the Population III stars. Moreover, our findings in abundance fitting procedure suggest that a better coverage of species before Zinc will significantly improve the performance of progenitor derivation. More attention should be paid to expand the coverage on light elements for future high-resolution spectroscopic observations on metal-poor stars. 
	
	\begin{acknowledgments}
        This study is supported by the National Natural Science Foundation of China under grant Nos. 11988101 and 12222305, National Key R\&D Program of China Nos. 2019YFA0405500. 
   % and CAS Project for Young Scientists in Basic Research grant No. YSBR-062. 
        H.L. and Q.X. acknowledge support from the Strategic Priority Research Program of Chinese Academy of Sciences grant No. XDB34020205, the Youth Innovation Promotion Association of the CAS (id. Y202017 and 2020058), and the science research grants from the China Manned Space Project. 
	\end{acknowledgments}
	
	\appendix
	\section{Fitting parameters of IMF} \label{sec:param_value}
	\begin{table*}[ht]
		\caption{The fitting parameters of \ac{IMF} and corresponding uncertainties with power-law and modified power-law functions}
		\label{tab:bestfit_params}
		\begin{tabular}[c]{c|c|cc|ccc}
			\hline \hline
			\multirow{2}{*}{\parbox{2cm}{Enrichment\\scenario}} & \multirow{2}{*}{\parbox{2cm}{Explodability\\restriction}} & \multicolumn{2}{c|}{Power-law} & \multicolumn{3}{c}{Modified power-law} \\
			&  & $\alpha_{m,\mathrm{CC}}$ & $\alpha_{m,\mathrm{PI}}$ & $\alpha_{m,\mathrm{CC}}$ & $\alpha_e$ & $\alpha_{m,\mathrm{PI}}$ \\ \hline
			\multirow{3}{*}{Mono} & ---  & $3.39\pm0.04$ & --- & --- & --- & --- \\
			& $2.35\ M_\odot$ & $3.58\pm0.04$ & --- & $-0.14^{+0.08}_{-0.07}$ & $2.60\pm0.04$ & --- \\
			& $1.94\ M_\odot$ & $3.75\pm0.04$ & --- & $-0.72^{+0.08}_{-0.10}$ & $2.64\pm0.04$ & --- \\ \hline
			\multirow{3}{*}{Duo} & --- & $4.63^{+0.31}_{-0.29}$ & $5.29^{+0.44}_{-0.52}$ 
                    & --- & --- & --- \\
			& $2.35\ M_\odot$ & $5.79^{+0.42}_{-0.39}$ & $5.18^{+0.46}_{-0.56}$ 
                    & $-0.69\pm0.74$ & $3.64^{+0.37}_{-0.35}$ & $5.22^{+0.56}_{-0.53}$ \\
			& $1.94\ M_\odot$ & $6.01^{+0.44}_{-0.41}$ & $5.20^{+0.47}_{-0.53}$ 
                    & $-0.32\pm0.74$ & $3.17\pm{0.30}$ & $5.33^{+0.59}_{-0.56}$
			%$\alpha_{m,\mathrm{CC}}=-0.24^{+0.77}_{-0.73}$ ($-0.66^{+0.71}_{-0.67}$), $\alpha_{e}=3.13\pm{0.30}$ ($3.60^{+0.35}_{-0.34}$) and $\alpha_{m,\mathrm{PI}}=5.33^{+0.58}_{-0.60}$ ($5.24^{+0.61}_{-0.57}$)
		\end{tabular}
	\end{table*}
	
	\bibliography{ExportDirectory}{}
	\bibliographystyle{aasjournal}
	
\end{document}